\documentclass[preprintnumbers]{revtex4}

\usepackage{graphicx}
\def\ie{{\it i.e.}}

\def\mpl{\ifmmode \overline M_{Pl}\else $\overline M_{Pl}$\fi}
\def\to{\rightarrow}
\newskip\zatskip \zatskip=0pt plus0pt minus0pt
\def\matth{\mathsurround=0pt}

\def\atversim#1#2{\lower0.7ex\vbox{\baselineskip\zatskip\lineskip\zatskip
  \lineskiplimit 0pt\ialign{$\matth#1\hfil##\hfil$\crcr#2\crcr\sim\crcr}}}

\def\be{\begin{equation}}
\def\ee{\end{equation}}
\def\bea{\begin{eqnarray}}
\def\eea{\end{eqnarray}}

\def\Re{{\cal R \mskip-4mu \lower.1ex \hbox{\it e}\,}}
\def\Im{{\cal I \mskip-5mu \lower.1ex \hbox{\it m}\,}}

\def\to{\rightarrow}
\def\slash{\not\!}
\def\inpb{\ifmmode {\rm pb}^{-1}\else ${\rm pb}^{-1}$\fi}
\def\infb{\ifmmode {\rm fb}^{-1}\else ${\rm fb}^{-1}$\fi}
\def\epem{\ifmmode e^+e^-\else $e^+e^-$\fi}
\def\sr{\ifmmode \tilde e_R\else $\tilde e_R$\fi}
\def\sl{\ifmmode \tilde e_L\else $\tilde e_L$\fi}

\def\lamt{\ifmmode \tilde\lambda\else $\tilde\lambda$\fi}
\def\shat{\ifmmode \hat s\else $\hat s$\fi}
\def\that{\ifmmode \hat t\else $\hat t$\fi}
\def\uhat{\ifmmode \hat u\else $\hat u$\fi}

\begin{document}
\bibliographystyle{revtex}
\rightline{\vbox{\halign{&#\hfil\cr
&SLAC-PUB-9608\cr
&December 2002\cr}}}
\vspace{2in} 
\begin{center}

{\Large Phenomenology of Supersymmetric Large Extra Dimensions
}
\medskip
\vspace{0.5in}

\normalsize 
{\large {JoAnne L. Hewett and Darius Sadri}}\\
\vskip .3cm
Stanford Linear Accelerator Center \\
Stanford CA 94309, USA\\
\vskip .3cm

\end{center} 
\vspace{1in}

We study the phenomenology of a supersymmetric bulk in
the scenario of large extra dimensions.
The virtual exchange of gravitino KK states in selectron pair production
in polarized \epem\ collisions is examined.  The
leading order operator for this exchange is dimension six, in
contrast to that of graviton KK exchange which induces a
dimension eight operator at lowest order.  Some kinematic
distributions for selectron production are presented. These processes
yield an enormous sensitivity to the fundamental higher dimensional Planck
scale.

\vskip1.5in
\noindent{Presented at the {\it The 10th International Conference
on Supersymmetry and Unification of Fundamental Interactions (SUSY02)}, 
Hamburg, Germany 17-23 June 2002;
and the {\it International Workshop on Linear Colliders (LCWS 2002)},
Jeju Island, Korea, 26-30 August 2002.}

\pagebreak

\title{Phenomenology of Supersymmetric Large Extra Dimensions}

\author{JoAnne L. Hewett and Darius Sadri}

\email[]{hewett@slac.stanford.edu,darius@stanford.edu}
\affiliation{Stanford Linear Accelerator Center, 
Stanford University, Stanford, California 94309 USA}


\begin{abstract}
We study the phenomenology of a supersymmetric bulk in
the scenario of large extra dimensions.
The virtual exchange of gravitino KK states in selectron pair production
in polarized \epem\ collisions is examined.  The
leading order operator for this exchange is dimension six, in
contrast to that of graviton KK exchange which induces a
dimension eight operator at lowest order.  Some kinematic
distributions for selectron production are presented. These processes
yield an enormous sensitivity to the fundamental higher dimensional Planck
scale.

\end{abstract}

\maketitle

One might wonder whether supersymmetry plays a role in
the recently proposed ADD scenario \cite{Arkani-Hamed:1998rs} of
large extra dimensions.
Clearly, bulk supersymmetry is not in conflict with the
basic assumptions of the model.  In fact, various reasons exist for believing
in a supersymmetric bulk,
not least of which is the motivation of string theory.
D-branes of string theory provide a
natural mechanism for the confinement of SM fields.
If string theory is the ultimate theory of nature then the proposal of ADD
might be embedded within it with a
bulk supporting a supersymmetric gravitational action, with
supersymmetry serving as a mechanism for stabilizing the bulk radii.
In \cite{sadri}, we investigated the consequences of a supersymmetric
bulk in the ADD scenario.  If bulk supersymmetry
remains unbroken away from the brane,
then it is natural to ask what happens
to the superpartners of the bulk gravitons, the gravitinos.
The bulk gravitinos must also expand into a Kaluza-Klein (KK) tower of states
and induce experimental signatures.

We focus on the effects of the virtual exchange of
the bulk gravitino and graviton KK tower states in the process
$e^+e^-\to\tilde e^+\tilde e^-$
at a high energy Linear Collider (LC).  This process is well-known
as a benchmark for
collider supersymmetry studies \cite{Baer:1988kx},
as the use of incoming polarized beams
enables one to disentangle the neutralino sector and determine the
degree of mixing between the various pure gaugino states.
The effects of the virtual exchange of a graviton KK tower
in selectron pair production has been examined in \cite{tgr} for
the case of non-supersymmetric large extra dimensions.
The introduction of gravitino KK exchange greatly alters the
phenomenology of this process by modifying the angular distributions
and by substantially
increasing the magnitude of the cross section.  We find that the
leading order behavior for this
process is given by a dimension-6 operator, in contrast to the
dimension-8 operator corresponding to graviton KK exchange.  This yields
a tremendous sensitivity to the existence of a supersymmetric bulk,
resulting in a search reach for the ultraviolet cut-off of the theory
of order $20-25\times\sqrt s$.

We assume that the
Standard Model fields are confined to a 3-brane.
In string theory, D-branes are extended objects on which open strings
terminate \cite{Polchinski:1996na}.
Only closed strings can propagate far away from
the D-brane, which on a ten dimensional background is
described locally by a type II string theory whose spectrum
contains two gravitinos (their vertex operators carry one vector and
one spinor index).
The D-brane introduces open string boundary conditions which are
invariant under just one supersymmetry, so only a linear
combination of the two original supersymmetries survives for the open
strings. Since open and closed strings
couple to each other, the D-brane breaks the original $N=2$
supersymmetry down to $N=1$. The low energy effective theory will then be a
$D=10, N=1$ supergravity theory with a single Majorana-Weyl gravitino
which couples to a
conserved space-time supercurrent.

We take the vacuum of space-time to be of the form
$M^{4} \times T^{6}$, where $M^{4}$ is four dimensional
Minkowski space-time and $T^{6} = S^{1} \times \ldots \times S^{1}$,
the direct product of six dimensions each
compactified on a circle, insuring four
dimensional Poincar\'{e} invariance.  For simplicity, we take a common
radius of compactification $R_c$ for all extra dimensions.
The gravitino kinetic term of the Lagrangian is
\begin{equation} \label{RS:action}
    E^{-1} \mathcal{L} \: = \:
         \frac{i}{2} \: \bar\Psi_{\hat{\mu}} \:
         \Gamma^{\hat{\mu}\hat{\nu}\hat{\rho}}
         \: \nabla_{\hat{\nu}}
         \Psi_{\hat{\rho}} \,,
\end{equation}
with $E$ being the determinant of the vielbein in ten dimensions,
$\Gamma^{\hat{\mu}\hat{\nu}\hat{\rho}}$ the antisymmetric product of three
$\Gamma$ matrices,
and $\Psi_{\hat{\mu}}$ a Majorana-Weyl vector-spinor.
Here the hatted indices range over all dimensions of the space-time.

The explicit derivation of an effective four dimensional action 
is presented in \cite{sadri}.  Here we
present a summary of the results. The 4-d effective Lagrangian can be
written as
\begin{equation} \label{effective:Lagrangian:KK:diag}
  e^{-1} \: \mathcal{L}_{eff}^{\vec{s}} ( x ) \: = \:
  \sum_{j=1}^{4}
  \left\{
  \frac{i}{2} \:
      \bar{\omega}_{m}^{\vec{s},j}(x)
      \gamma^{mnp} ( \partial_{n} \omega_{p}^{\vec{s},j} (x) )
                \: + \:
      i \: \bar{\omega}_{m}^{\vec{s},j}(x)
                \sigma^{mp} m_{\vec{s}}^j
                \omega_{p}^{\vec{s},j}(x)
  \right\}\,,
\end{equation}
with the mass given by
$m_{\vec{s}}^j = (-1)^j \frac{\sqrt{\vec{s} \cdot \vec{s}}}{R_{c}}$.
The fields associated with the negative mass eigenvalues can be redefined
to remove this sign, however, care must be taken with the Feynman rule for
the coupling of the gravitino to matter.
The sum in Eq. (\ref{effective:Lagrangian:KK:diag})
runs over the four Majorana vector-spinors.
We have applied the Majorana-Weyl condition in ten dimensions,
which yields four Majorana spinors after
the decomposition into four dimensions.
Generally, the masses of the four gravitinos at each Kaluza-Klein
level can be shifted by supersymmetry breaking effects on the brane.
When studying the phenomenology of such models,
we will assume that the $N=4$ supersymmetry
is broken at scales near the fundamental scale $M_D$, with only $N=1$
supersymmetry surviving down to the electroweak scale.
The phenomenological contributions from the heavy gravitinos associated
with the breaking of the extended supersymmetry
near the fundamental scale will be highly suppressed,
due to the large mass of these individual excitations.

In \cite{sadri},
we derive the coupling of fermions and scalars to the gravitinos.
The term coupling scalars, spinors and gravitinos,
and minimally coupled to gravity, is
\begin{equation}
  \mathcal{L}_I \: = \:
  - \frac{\kappa}{\sqrt{2}} | e |
  \left\{
    \left( \partial_\mu \Phi_L \right) \bar{\Omega}_\nu
    \gamma^\mu \gamma^\nu \psi_L \: + \:
    \left( \partial_\mu \Phi_R \right) \bar{\Omega}_\nu
    \gamma^\mu \gamma^\nu \psi_R
  \right\}\, \: + \: h.c. \,,
\end{equation}
with $\bar{\Omega}_\nu$ a Majorana vector-spinor.
Expanding $|e|$ to leading order in the vierbein yields the
appropriate Feynman rules.

There are numerous tree level processes contributing to selectron pair
production in the presence of a supersymmetric bulk.
In addition to the standard
$\gamma,Z$ s-channel exchange and $\tilde B^0,\tilde W^0$ t-channel
contributions present in the MSSM, we now have contributions arising from
the s-channel exchange of the bulk graviton KK tower and the
t-channel exchange of the bulk gravitino KK tower.
There are no u-channel contributions due to the non-identical final
states.  The contributions from neutral higgsino states
are negligible due to the smallness of the Yukawa coupling.
The diagrammatic contributions to the individual
scattering processes for left- and right-handed selectron production with
initial polarized electron beams are summarized in Table \ref{diag}.
Note that the $\tilde W^0$ exchange only contributes to the process
$e^-_Le^+\to\sl^-\sl^+$, and that the t-channel gravitino and 
the $\tilde B^0$
contributions are isolated in the reaction 
$e^-_{L,R}e^+\to\tilde e^-_{L,R}\tilde e^+_{R,L}$.

\begin{table}
\centering
\begin{tabular}{|c|c|c|c|c|} \hline\hline
 & $\tilde e^-_L\tilde e^+_L$ & $\tilde e^-_R\tilde e^+_L$ &
   $\tilde e^-_L\tilde e^+_R $ & $\tilde e^-_R\tilde e^+_R$ \\ \hline
$e^-_Le^+$ & s-channel $\gamma\,, Z\,, G_n$ & & & s-channel $\gamma\,, Z\,,
  G_n$ \\
  & t-channel $\tilde W\,, \tilde B\,, \Psi_n$ & & t-channel $\tilde B\,,
    \Psi_n$ & \\ \hline
$e^-_Re^+$ & s-channel $\gamma\,, Z\,, G_n$ & & & s-channel $\gamma\,, Z\,,
G_n$ \\
  & & t-channel $\tilde B\,, \Psi_n$ & & t-channel $\tilde B\,, \Psi_n$ \\
\hline\hline
\end{tabular}
\caption{The diagrammatic contributions to individual scattering processes
for polarized electron beams.  A blank box indicates that there are no
contributions for that polarization configuration.}
\label{diag}
\end{table}

The unpolarized matrix element for the case of massive gravitino KK
exchanges is
\begin{equation}
{\cal M}={\frac{\kappa^2}{2}}\sum_{\vec n} {\frac{k_1^\nu k_2^\rho}
{t-m^2_{\vec n}}}\bar e(p_1)\gamma_\mu\gamma_\nu P^{\vec n,\mu\tau}
\gamma_\rho\gamma_\tau e(p_2)\,,
\end{equation}
where the sum extends over the gravitino KK modes and
$\kappa=\sqrt{8\pi G_N} =\bar M_P^{-1}$
is the reduced Planck scale.   $P^{\vec n,\mu\tau}$ represents the
numerator of the propagator for a Rarita-Schwinger field of mass
$m_{\vec n}$ and is given by
\begin{equation} \label{propagator}
  P^{\vec{n}, \mu \nu} \: = \:
  i \: \left( \slash{k} + m_{\vec{n}} \right)
  \left( \frac{k^{\mu} k^{\nu}}{m_{\vec{n}}^2} - \eta^{\mu \nu} \right)
  - \frac{i}{3}
  \left( \gamma^{\mu} + \frac{k^{\mu}}{m_{\vec{n}}} \right)
  \left( \slash{k} - m_{\vec{n}} \right)
  \left( \gamma^{\nu} + \frac{k^{\nu}}{m_{\vec{n}}} \right)\,.
\end{equation}
The mass splitting between the evenly spaced bulk gravitino KK
excitations is given by
$1/R_c$, which lies in the range $10^{-4}$ eV to few MeV for $\delta=2$ to
6 assuming $M_D\sim 1$ TeV; their number density is thus large at
collider energies.
The sum over the KK states can then be approximated
by an integral
which is log divergent for $\delta=2$ and power divergent for
$\delta>2$.  We employ a cut-off to regulate these ultraviolet
divergences, with the cut-off being set to $\Lambda_c$, which in
general is different from $M_D$, to
account for the uncertainties from the unknown ultraviolet physics.
This approach is the most model independent and is that
generally used in the case of virtual graviton exchange \cite{Hewett:1998sn}.
In practice,
the integral over the gravitino KK states is more
complicated than that in the case with spin-2 gravitons due to the 
dependence of the gravitino propagator on $m_{\vec n}$.
We find that the leading order term for $\sqrt{|t|}\ll\Lambda_c$
results in
the replacement (in the case of $\delta=6$)
\begin{equation}
{\frac{\kappa^2}{2}}\sum_{\vec n} {\frac{P^{\vec n,\mu\tau}}{t-m^2_{\vec n}}}
\to {\frac{-i8\pi}{5\Lambda_c^3}}\left( \eta^{\mu\nu}-{\frac{1}{3}}
\gamma^\mu\gamma^\nu\right) \,
\end{equation}
in the matrix element; the structure of the summed gravitino propagator
is thus altered from that of a single massive state.
Hence the leading order behavior for gravitino KK exchange results in
a dimension-6 operator!  This is in stark contrast to graviton KK
exchange which yields a dimension-8 operator at leading order.
We thus expect an increased sensitivity to the scale $\Lambda_c$ in
the case of a supersymmetric bulk.

In order to perform a numerical analysis of this process, we need to specify
a concrete supersymmetric model.  We choose that of Gauge 
Mediated Supersymmetry
Breaking (GSMB) as it naturally contains a light zero-mode gravitino.
We specify a sample set of input parameters at the messenger scale, where the
supersymmetry breaking is mediated via the messenger sector, and
use the Renormalization Group Equations (RGE) to obtain the low-energy
sparticle spectrum.  We choose two sets of sample input parameters
describing the messenger sector which are consistent with our model.
The RGE evolution of these parameter sets 
results in the sparticle spectrum
\begin{eqnarray}
{\rm Set~I} : &  m_{\tilde e_L} & =217.0\, {\rm GeV},\quad
m_{\tilde e_R}=108.0\, {\rm GeV}, \quad
\chi^0_i=(76.5,\, 141.5,\, 337.0,\, 367.0)\, {\rm GeV} \,,\nonumber \\
{\rm Set~II}: &  m_{\tilde e_L} & =210.5\, {\rm GeV},\quad
m_{\tilde e_R}=104.5\, {\rm GeV}, \quad
\chi^0_i=(110.5,\, 209.6,\, 322.5,\, 324.0)\, {\rm GeV}  \,,
\nonumber
\end{eqnarray}
where $\chi^0_i$ with $i=1,4$ corresponds to the four mixed neutralino
states.
The first set of parameters yields a bino-like state for the lightest
neutralino, whereas
the second set results in a Higgsino-like state for $\chi^0_1$.
These input parameters were selected in order to obtain a sparticle spectrum
which is kinematically accessible to the Linear Collider; our
results are essentially insensitive to the exact details of the spectrum.

\begin{figure}[htbp]
\centerline{
\includegraphics[width=5.1cm,angle=90]{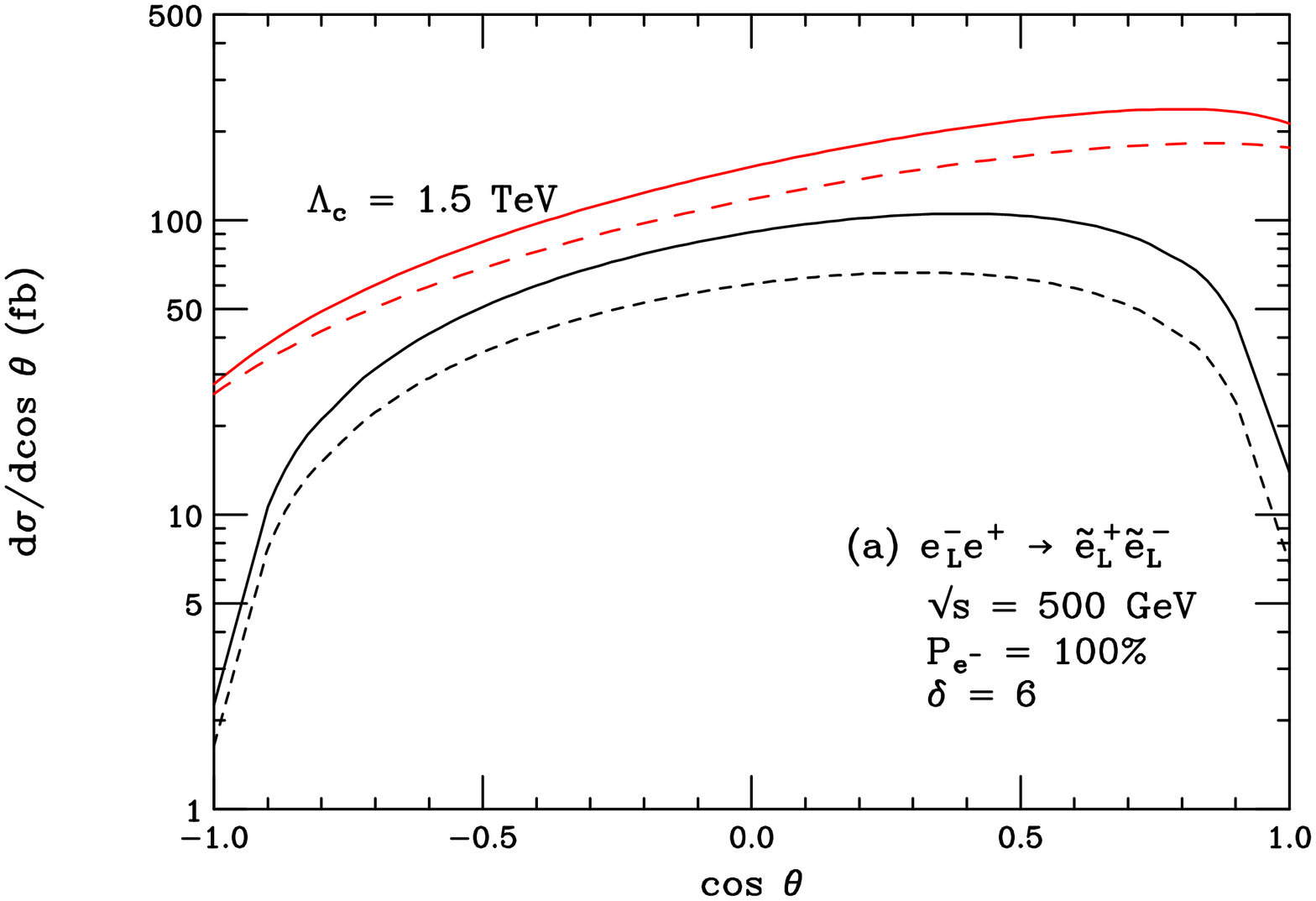}
\hspace*{5mm}
\includegraphics[width=5.1cm,angle=90]{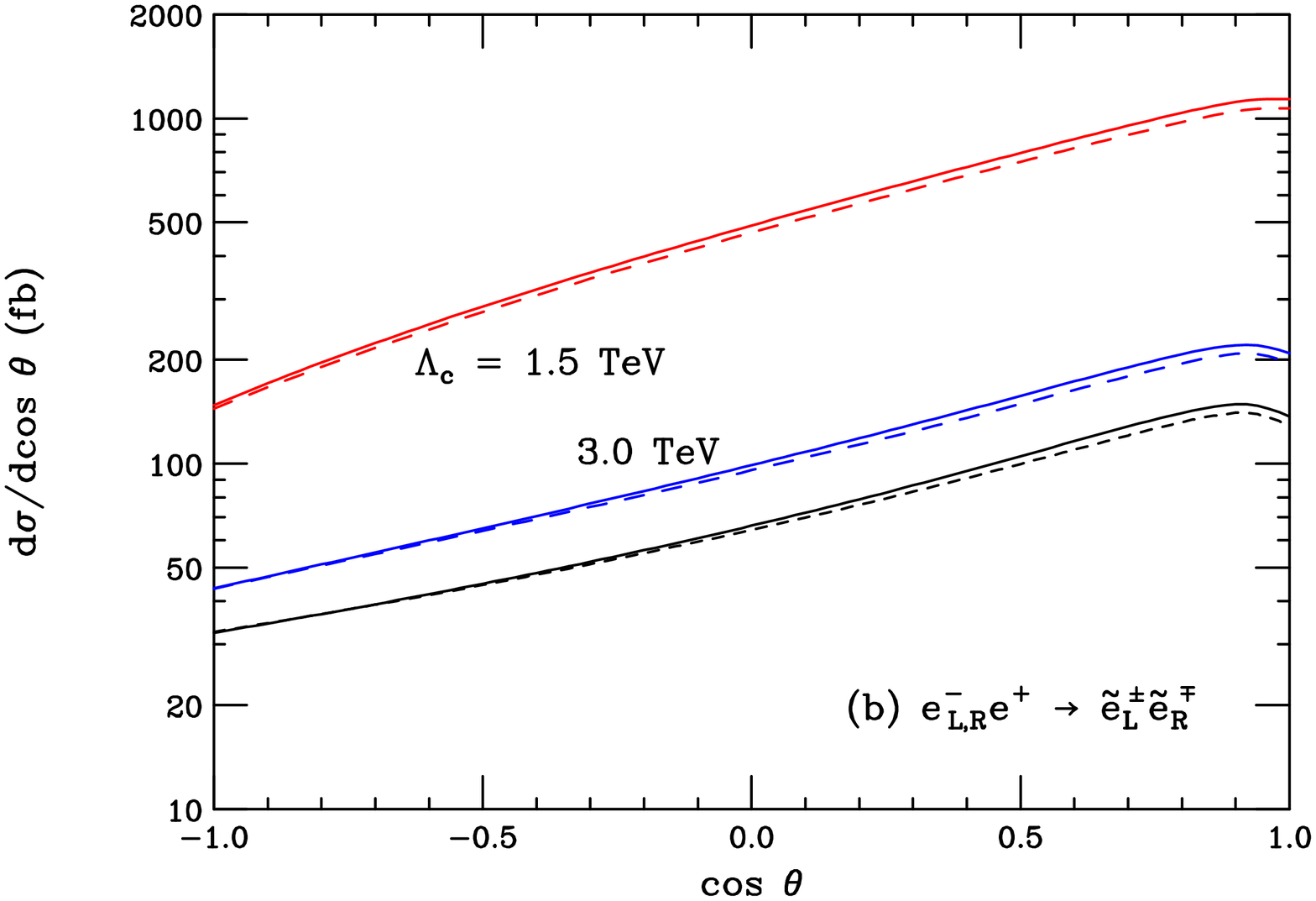}}
\vspace*{0.1cm}
\centerline{
\includegraphics[width=5.1cm,angle=90]{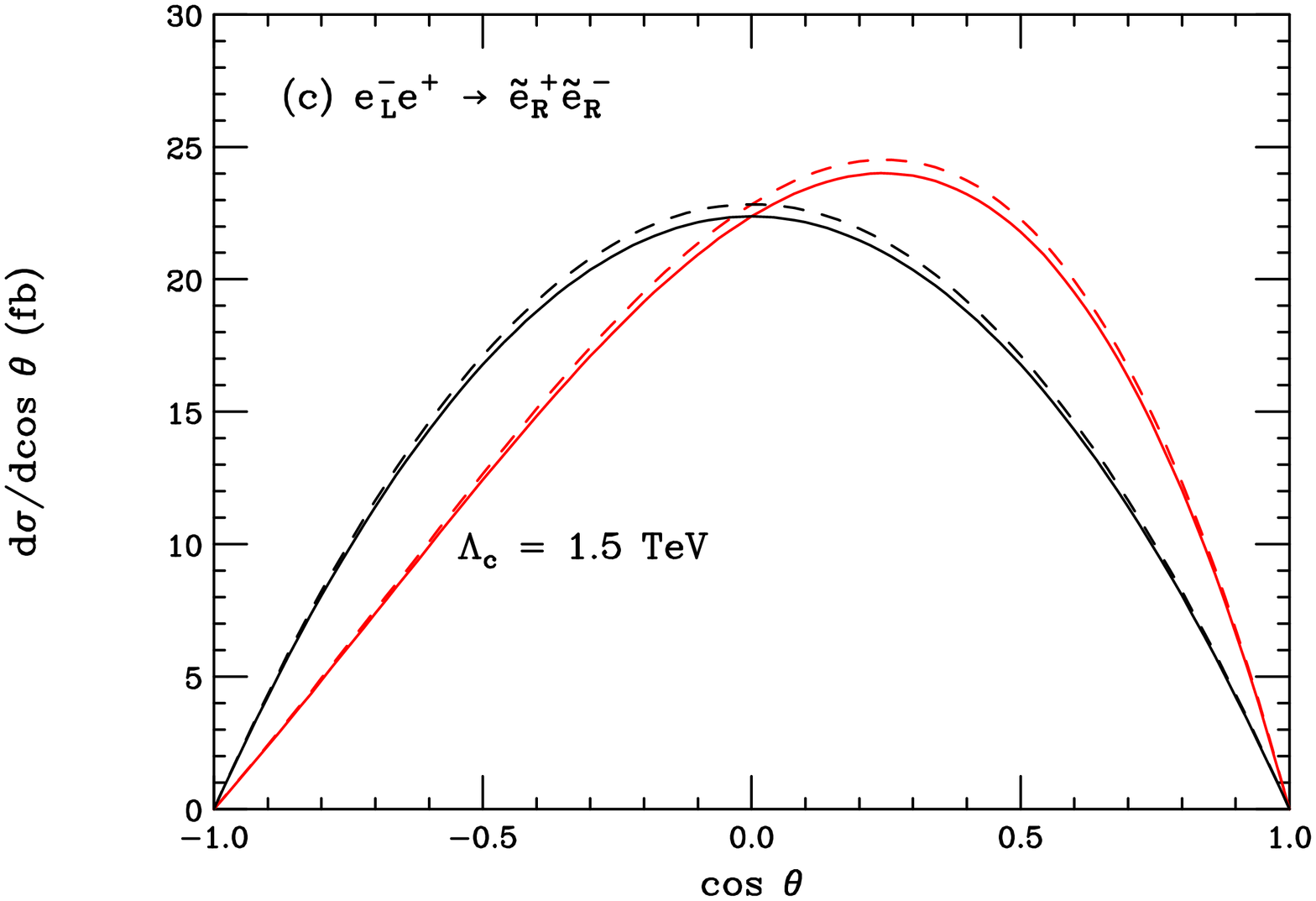}
\hspace*{5mm}
\includegraphics[width=5.1cm,angle=90]{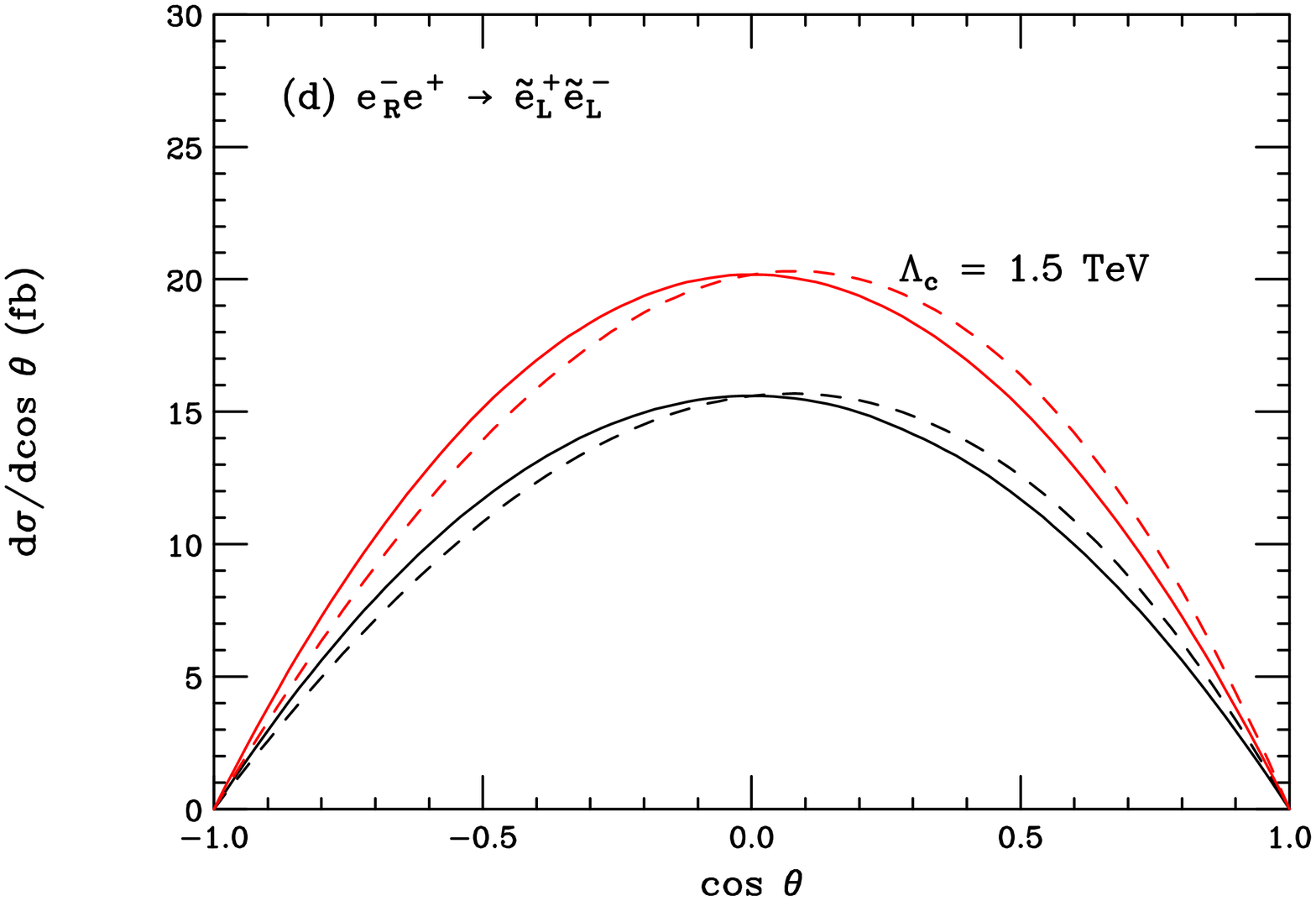}}
\vspace*{0.1cm}
\centerline{
\includegraphics[width=5.1cm,angle=90]{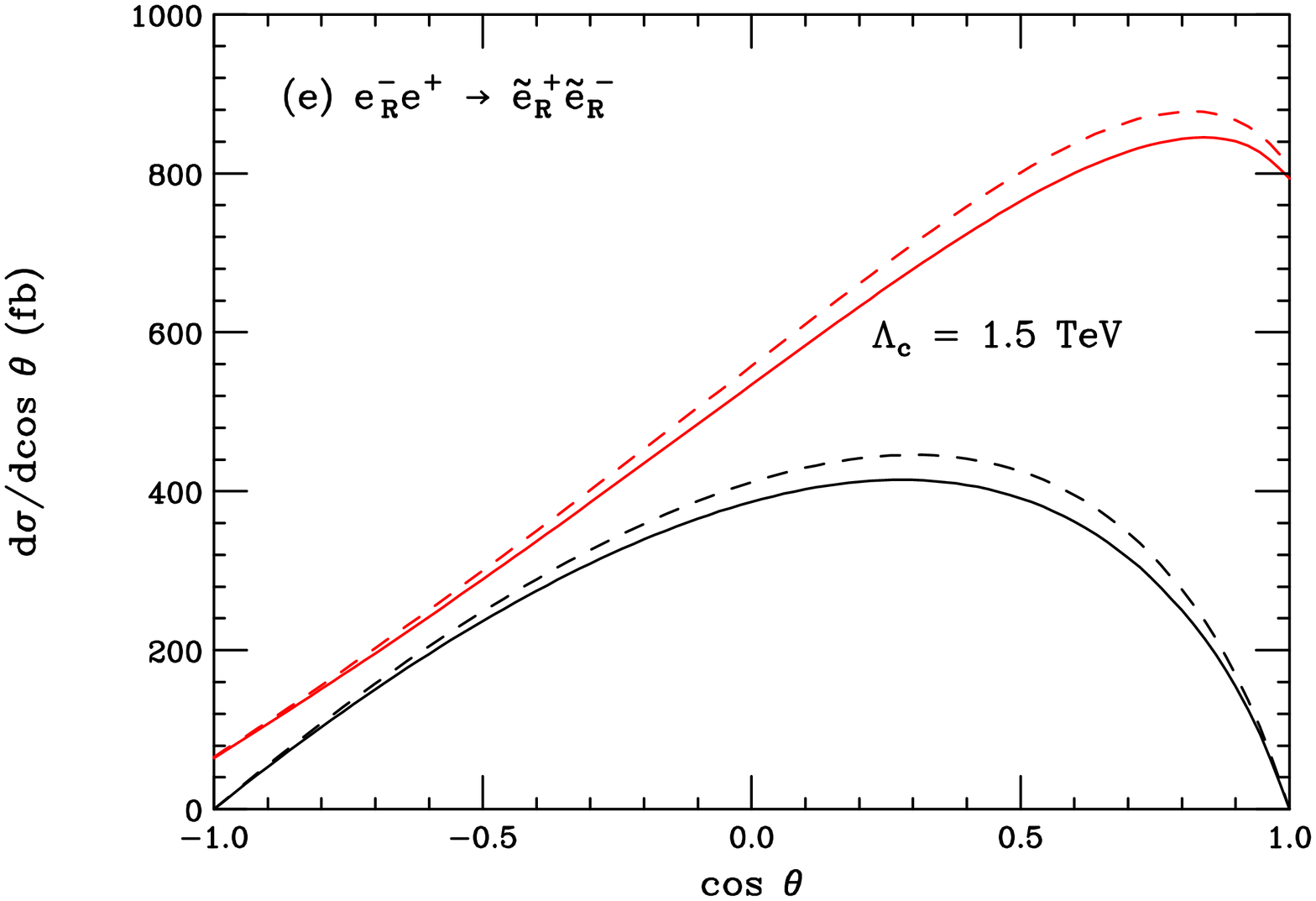}}
\vspace*{0.1cm}
\caption{Angular distributions for each helicity configuration with
supersymmetric bulk contributions for $\Lambda_c=1.5$ TeV (top
curves), and for  the $D=4$ supersymmetric models (bottom curves).
The solid (dashed) curves correspond to a bino-like (Higgsino-like)
composition of the lightest neutralino.}
\label{8782_fig5}
\end{figure}

Figure \ref{8782_fig5} shows the
angular distributions with $100\%$ electron beam polarization
for each helicity configuration listed in Table \ref{diag} for the two
sets of parameters discussed above, with and without the contributions
from supersymmetric extra dimensions.  In each case, the solid curve 
corresponds to the bino-like case and the dashed curve
represents the Higgsino-like scenario.  The top set of curves are those
for a supersymmetric bulk with $\Lambda_c=1.5$ TeV, while the
bottom set corresponds to our two $D=4$ supersymmetric models,
\ie, without the graviton and gravitino KK contributions.  
We note that the $D=4$ results (i.e. MSSM) agree with  those in the
literature \cite{Baer:1988kx}.  We see from the figure that in the 
process where the gravitino contributions are dominant, 
$e^-_{L,R}e^+\to\sl^\pm\sr^\mp$, there is little
difference in the shape or magnitude between the two $\chi^0_1$
compositions.  The use of
selectron pair production in polarized \epem\ collisions as a means of
determining the composition of the lightest neutralino is thus made
more difficult in the scenario with supersymmetric large extra dimensions.
In what follows, we present results only
for the bino-like $\chi^0_1$ as a sample case; our conclusions will not
be dependent on the assumptions of the composition of the lightest
neutralino.

\begin{figure}[htbp]
\centerline{
\includegraphics[width=6cm,angle=90]{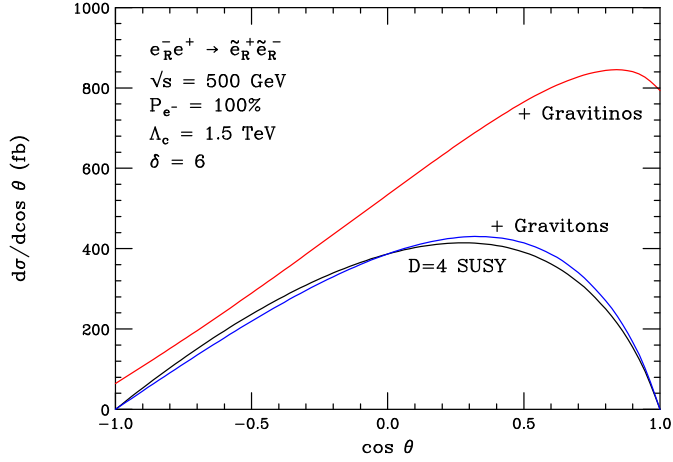}}
\vspace*{0.1cm}
\caption{The angular distribution for $e^-_Re^+\to\sr^+\sr^-$ from
the $D=4$ supersymmetric model I, plus the addition of bulk graviton
KK tower exchange, and with bulk gravitino KK tower exchange,
corresponding to the bottom, middle, and top curves, respectively.}
\label{8782_fig3}
\end{figure}

Figure \ref{8782_fig3} shows the angular distribution for
the process $e^-_Re^+\to\sr^+\sr^-$ with $\sqrt s=500$ GeV assuming
100\% polarization of the electron beam,
detailing the effects of each class of contributions
to selectron pair production.
The bottom curve represents the full contributions
(s- and t-channel) from the 4-dimensional standard
gauge-mediated supersymmetric model discussed above in the case where 
the $\chi^0_1$ is bino-like, corresponding to parameter set I.
The middle curve displays the effects of adding only 
the s-channel contributions of the bulk graviton KK tower
in the scenario of a non-supersymmetric bulk with $\Lambda_c=1.5$ TeV.
We see that there
is little difference in the distribution between the $D=4$ supersymmetric
case and with the addition of the
graviton KK tower, in either shape or magnitude.  It would hence be 
difficult to disentangle the effects of graviton exchange from an accurate
measurement of the underlying supersymmetric parameters using this
process alone.  The top curve corresponds to the full set of contributions
from a supersymmetric bulk, \ie, our standard supersymmetric model plus
KK graviton and KK gravitino tower exchange
for the case of six extra dimensions with
$\Lambda_c=1.5$ TeV.  Here we see that the exchange of bulk gravitino 
KK states yields a large enhancement in the cross section and a substantial
shift in the shape of the angular distribution, particularly at forward
angles, even for $\Lambda_c=3\sqrt s$.
This provides a dramatic signal for a supersymmetric bulk!

\begin{figure}[htbp]
\centerline{
\includegraphics[width=5.1cm,angle=90]{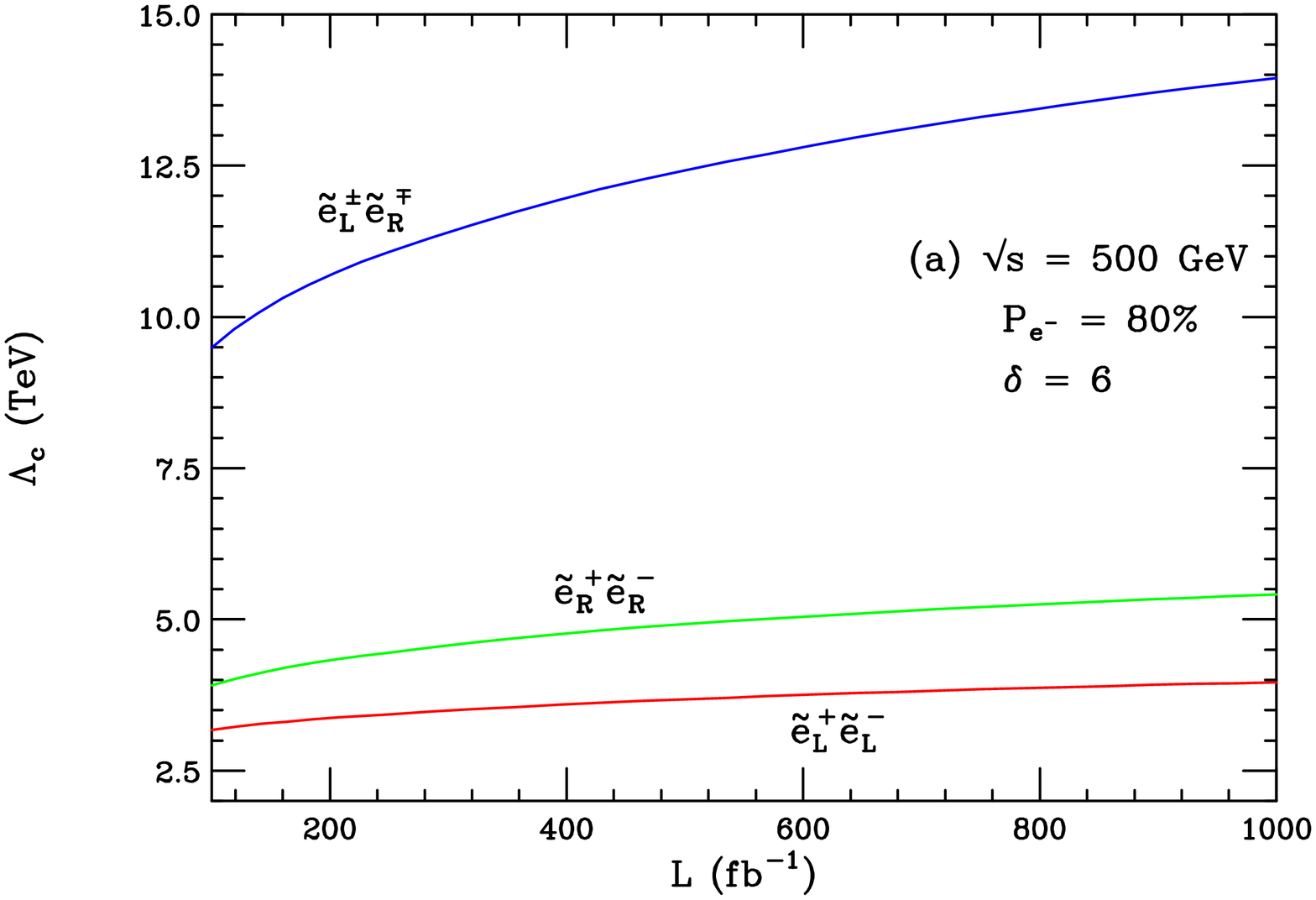}
\hspace*{5mm}
\includegraphics[width=5.1cm,angle=90]{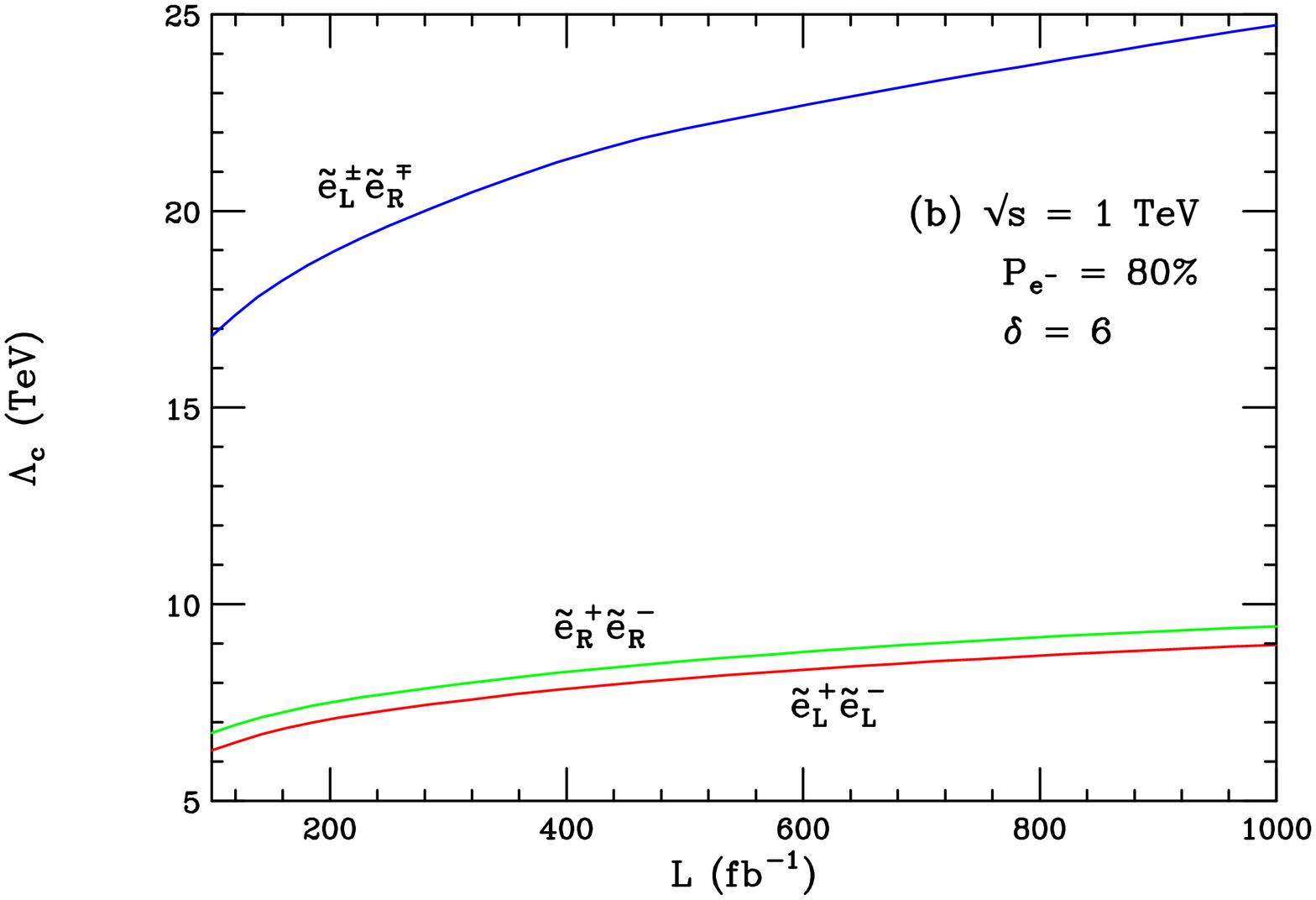}}
\vspace*{0.1cm}
\caption{95\% C.L. search reach for $\Lambda_c$ in each production
channel as a function of
integrated luminosity for $\sqrt s= 0.5$ and 1.0 TeV.}
\label{8782_fig9}
\end{figure}

We now compute the potential sensitivity to the cut-off scale from 
selectron pair production using our sample case with a bino-like 
lightest neutralino state.
We employ the 
usual $\chi^2$ procedure,
including statistical errors only.  We sum over both initial left-
and right-handed electron polarization states, assuming $P_{e^-}=80\%$.
The resulting 95\% C.L. search for $\Lambda_c$ from each final
state, $\sl^+\sl^-\,, \sr^+\sr^-$, and $\sl^\pm\sr^\mp$, is given as a
function
of integrated luminosity in Fig. \ref {8782_fig9} for $\sqrt s = 0.5$ and 1.0
TeV.  We see that for 500 \infb\ of integrated luminosity, corresponding
to design values, the search reach in the left- and right-handed selectron
pair production channels is given roughly by $\Lambda_c\simeq 6-10
\times\sqrt s$, which is essentially what is achievable for bulk graviton KK
exchange  in the reaction $e^+e^-\to f\bar f$ \cite{Hewett:1998sn}.
However, the $\sl^\pm\sr^\mp$ production channel yields an enormous
search capability with a 95\% C.L. sensitivity to $\Lambda_c$ of order
$25\times\sqrt s$ for design luminosity.  This process thus has the potential
to either discover a supersymmetric bulk, or eliminate the possibility of
supersymmetric large extra dimensions as being relevant to the hierarchy
problem.  We stress that there is nothing special about our choice of
supersymmetric parameters; our results will hold as long as selectrons 
are kinematically accessible to high energy \epem\ colliders.
We conclude that selectron pair production provides a very powerful
tool in searching for a supersymmetric bulk.

In summary, we have examined the phenomenological consequences of a 
supersymmetric bulk in the scenario of large extra dimensions.  We assumed
that supersymmetry is unbroken in the bulk, with gravitons and gravitinos
being free to propagate throughout the higher dimensional space, 
and that the SM and MSSM gauge and matter
fields are confined to a 3-brane.  Motivated by string theory, we
worked in the framework of $D=10$
supergravity, and found that the KK reduction of the bulk gravitinos
yields four Majorana spinors in four dimensions.  We then assumed that
the residual $N=4$ supersymmetry is broken near the fundamental scale 
$M_D$, with only $N=1$ supersymmetry surviving at the
electroweak scale.  

Starting with the $D=10$ action for this scenario, we expanded the
bulk gravitino into a KK tower of states, and determined the 4-d action
for the spin-3/2 KK excitations.  We then presented
the coupling of the bulk gravitino KK states to fermions and their
scalar partners on the brane.  We applied these results to a
phenomenological analysis by examining
the effects of virtual exchange of the  gravitino
KK tower in superparticle pair production.  We focused on the
reaction $\epem\to\tilde e^+\tilde e^-$ as this process is a
benchmark for collider supersymmetry studies.  Our numerical
analysis was performed in the framework of gauge mediated
supersymmetry breaking as it naturally affords a light
zero-mode gravitino.  However, our results do not depend on
the specifics of this particular model, with the exception of
the existence of a light zero-mode gravitino state.

Performing the sum over the KK propagators, we found that
the leading order contribution to this process arises from a
dimension-6 operator, and is independent of the zero-mode
mass.  This is in stark contrast to the virtual exchange of
spin-2 graviton KK states, which yields a dimension-8 operator
at leading order.  We thus found that the gravitino KK
contributions  substantially alter
the production rates and angular distributions for 
selectron pair production, and may essentially be isolated
in the $\sl^\pm\sr^\mp$ channel.  The resulting sensitivity
to the cut-off scale is tremendous, being of order
$20-25\times\sqrt s$.

We expect that the virtual exchange of gravitino KK
states in hadronic collisions will have somewhat less of an effect
in squark and gluino pair production than what we have found 
here.  The reason is that these
processes are initiated by both quark annihilation and gluon
fusion sub-processes, only one of which will be sensitive to tree-level
gravitino exchange for a given production channel.  The
sensitivity to the cut-off scale will then depend on the
relative weighting of the quark and gluon initial states.
In addition, t-channel gravitino
contributions will only be numerically relevant
for up- and down-squark production
due to flavor conservation;
hence their effect will be diluted by the production of the other
degenerate squark flavors and the relative weighting of the parton
densities.

We note also that virtual exchange of gravitino KK states
may also have a large effect on selectron pair production in
$e^-e^-$ collisions, which are tailor-made for  t-channel
Majorana exchanges.  High energy Linear Colliders thus provide
an excellent probe for the existence of  supersymmetric large
extra dimensions, and have the capability of discovering this
possibility or eliminating it as being relevant to the hierarchy
problem.

\begin{acknowledgments}

The authors are supported by the US Department of Energy, Contract
DE-AC03-76SF00515.

\end{acknowledgments}

\def\MPL #1 #2 #3 {Mod. Phys. Lett. {\bf#1},\ #2 (#3)}
\def\NPB #1 #2 #3 {Nucl. Phys. {\bf#1},\ #2 (#3)}
\def\PLB #1 #2 #3 {Phys. Lett. {\bf#1},\ #2 (#3)}
\def\PR #1 #2 #3 {Phys. Rep. {\bf#1},\ #2 (#3)}
\def\PRD #1 #2 #3 {Phys. Rev. {\bf#1},\ #2 (#3)}
\def\PRL #1 #2 #3 {Phys. Rev. Lett. {\bf#1},\ #2 (#3)}
\def\RMP #1 #2 #3 {Rev. Mod. Phys. {\bf#1},\ #2 (#3)}
\def\NIM #1 #2 #3 {Nuc. Inst. Meth. {\bf#1},\ #2 (#3)}
\def\ZPC #1 #2 #3 {Z. Phys. {\bf#1},\ #2 (#3)}
\def\EJPC #1 #2 #3 {E. Phys. J. {\bf#1},\ #2 (#3)}
\def\IJMP #1 #2 #3 {Int. J. Mod. Phys. {\bf#1},\ #2 (#3)}
\def\JHEP #1 #2 #3 {J. High En. Phys. {\bf#1},\ #2 (#3)}

\end{document}